%% file: EXO-12-050_temp.tex
\pdfoutput=1

\documentclass[11pt,twoside,a4paper,cmspaper,final,collab]{cms-tdr}
\def\svnVersion{289914}\def\cmsCernNo{2013-037}\def\cmsCernDate{\today}\def\cmsMessage{Published in Physics Letters B as \href{http://dx.doi.org/10.1016/j.physletb.2015.04.042}{\doi{10.1016/j.physletb.2015.04.042.}}}

\begin{document}\cmsNoteHeader{EXO-12-050}

\hyphenation{had-ron-i-za-tion}
\hyphenation{cal-or-i-me-ter}
\hyphenation{de-vices}
\RCS$Revision: 284702 $
\RCS$HeadURL: svn+ssh://svn.cern.ch/reps/tdr2/papers/EXO-12-050/trunk/EXO-12-050.tex $
\RCS$Id: EXO-12-050.tex 284702 2015-04-15 16:50:54Z apana $
\newlength\cmsFigWidth
\ifthenelse{\boolean{cms@external}}{\setlength\cmsFigWidth{0.99\columnwidth}}{\setlength\cmsFigWidth{0.7\textwidth}}
\ifthenelse{\boolean{cms@external}}{\providecommand{\cmsLeft}{top}}{\providecommand{\cmsLeft}{left}}
\ifthenelse{\boolean{cms@external}}{\providecommand{\cmsRight}{bottom}}{\providecommand{\cmsRight}{right}}

\providecommand{\nED}{\ensuremath{n_\mathrm{ED}}\xspace}
\providecommand{\CIJET} {{\textsc{cijet}}\xspace}
\providecommand{\CIDIJET} {{\textsc{cidijet}}\xspace}
\providecommand{\NLOJETPP} {{\textsc{NLOJet++}}\xspace}
\providecommand{\fastNLO} {{\textsc{fastNLO}}\xspace}
\providecommand{\fastjet} {{\textsc{FastJet}}\xspace}
\providecommand{\GeVsq}{\ensuremath{\GeV^2}\xspace}
\providecommand{\PYTHIAS} {{\textsc{pythia 6}}\xspace}
\providecommand{\PYTHIAE} {{\textsc{pythia 8}}\xspace}
\providecommand{\HERWIGpp} {{\textsc{herwig++}}\xspace}
\providecommand{\RooUnfold} {{\textsc{RooUnfold}}\xspace}
\providecommand{\ptmax}{\ensuremath{p_{\mathrm{T,max}}}\xspace}
\providecommand{\mur}{\ensuremath{\mu_R}\xspace}
\providecommand{\muf}{\ensuremath{\mu_F}\xspace}
\providecommand{\rbthm}{\rule[-2ex]{0ex}{5ex}}
\providecommand{\rbtrr}{\rule[-0.8ex]{0ex}{3.2ex}}
\providecommand{\chijj}{\ensuremath{\chi_\text{dijet}}\xspace}
\providecommand{\ystar}{\ensuremath{y^{\star}}\xspace}
\providecommand{\yboost}{\ensuremath{y_\text{boost}}\xspace}
\providecommand{\intlumi} {19.7\fbinv\xspace}
\providecommand{\mjj} {\ensuremath{M_{jj}}\xspace}
\providecommand{\avept}{\ensuremath{\langle p_\mathrm{T1,2}\rangle}\xspace}

\cmsNoteHeader{EXO-12-050} % This is over-written in the CMS environment: useful as preprint no. for export versions
\title{Search for quark contact interactions and extra spatial
dimensions using dijet angular distributions in proton-proton
collisions at $\sqrt{s} = 8$\TeV}

\date{\today}

\abstract{A search is presented for quark contact interactions and
  extra spatial dimensions in proton-proton collisions at $\sqrt{s} =
  8$\TeV using dijet angular distributions.  The search is based on a
  data set corresponding to an integrated luminosity of 19.7\fbinv collected by the CMS detector at the CERN LHC. Dijet
  angular distributions are found to be in agreement with the
  perturbative QCD predictions that include electroweak corrections.
  Limits on the contact interaction scale from a variety of models at
  next-to-leading order in QCD corrections are obtained.  A benchmark
  model in which only left-handed quarks participate is excluded up to
  a scale of 9.0 (11.7)\TeV for destructive (constructive)
  interference at 95\% confidence level.  Lower limits between 5.9
  and 8.4\TeV on the scale of virtual graviton exchange are
  extracted for the Arkani-Hamed--Dimopoulos--Dvali model of extra
  spatial dimensions.}

\hypersetup{%
pdfauthor={CMS Collaboration},%
pdftitle={Search for quark contact interactions and extra spatial
  dimensions using dijet angular distributions in proton-proton
  collisions at sqrt(s) = 8 TeV},%
pdfsubject={CMS},%
pdfkeywords={CMS, physics, QCD, electroweak corrections, contact interactions, extra dimensions}}

\maketitle
\section{Introduction}

High momentum-transfer proton-proton collisions at the CERN LHC
probe the dynamics of the underlying interaction at distances below
$10^{-19}\unit{m}$. Often these collisions produce a pair of jets (dijets) approximately
balanced in transverse momentum \pt. These
dijet events provide an ideal testing ground to probe the validity of perturbative
quantum chromodynamics and to search for new phenomena such as
quark compositeness or additional,
compactified spatial dimensions. A particularly suitable observable
for this purpose is the dijet angular distribution~\cite{UA1} expressed
in terms of $\chijj = \exp(\abs{y_1 -y_2})$, where $y_1$ and $y_2$ are the
rapidities of the two jets with the highest transverse momenta.
Rapidity is defined as $y = \ln\left[\left(E + p_z\right) / \left(E -
    p_z\right)\right]/2$ with $E$ being the jet energy and $p_z$ the
projection of the jet momentum onto the beam axis.
For the scattering of massless partons, \chijj is related to the polar scattering angle $\theta^{*}$ in the partonic center-of-mass (c.m.) frame by
$\chijj = (1 + \abs{\cos{\theta^*}}) / ( 1 - \abs{\cos{\theta^*}})$.
The choice of the variable $\chijj$ is motivated by the fact that for
Rutherford scattering the angular distribution is approximately independent of
$\chijj$.
In perturbative QCD the dijet angular distribution at small c.m. scattering angles
is approximately independent of the underlying partonic level process
and exhibits
behavior similar to Rutherford scattering, characteristic of spin-1 particle
exchange.
Signatures of new
physics (NP), such as quark contact interactions (CI) or virtual
exchange of Kaluza--Klein~\cite{Klein} excitations of the graviton,
that exhibit angular distributions that are more isotropic than those
predicted by QCD, could appear as an excess of events at low values of \chijj.

Models of quark compositeness~\cite{Terazawa:1980,Eichten:1983hw,Eichten:1984eu}
postulate interactions between quark constituents at a
characteristic scale $\Lambda$ that is much larger than the quark
masses. At energies well below $\Lambda$, these interactions can be
approximated by a CI characterized by a
four-fermion coupling. The effective Lagrangian for
flavor-diagonal color-singlet couplings between quarks can be written
as~\cite{Eichten:1983hw,Eichten:1984eu}:
\ifthenelse{\boolean{cms@external}}{
\begin{multline*}
  \mathcal{L}_{\cPq\cPq}=\frac{2\pi}{\Lambda^2} \big[
    \eta_{LL}({\overline{\cPq}}_{L}\gamma^{\mu}\cPq_{L})(\overline{\cPq}_{L}\gamma_{\mu}\cPq_{L})\\
    +\eta_{RR}(\overline{\cPq}_{R}\gamma^{\mu}\cPq_{R})(\overline{\cPq}_{R}\gamma_{\mu}\cPq_{R})
    +2\eta_{RL}(\overline{\cPq}_{R}\gamma^{\mu}\cPq_{R})(\overline{\cPq}_{L}\gamma_{\mu}\cPq_{L})
  \big],
\end{multline*}
}{
\begin{equation*}
  \mathcal{L}_{\cPq\cPq}=\frac{2\pi}{\Lambda^2} \left[
    \eta_{LL}({\overline{\cPq}}_{L}\gamma^{\mu}\cPq_{L})(\overline{\cPq}_{L}\gamma_{\mu}\cPq_{L})
    +\eta_{RR}(\overline{\cPq}_{R}\gamma^{\mu}\cPq_{R})(\overline{\cPq}_{R}\gamma_{\mu}\cPq_{R})
    +2\eta_{RL}(\overline{\cPq}_{R}\gamma^{\mu}\cPq_{R})(\overline{\cPq}_{L}\gamma_{\mu}\cPq_{L})
  \right],
\end{equation*}
}
where the subscripts $L$ and $R$ refer to the left and right chiral projections of
the quark fields respectively and $\eta_{LL}$, $\eta_{RR}$, and $\eta_{RL}$ are
taken to be
0, ${+}1$, or ${-}1$. The various combinations of
$\left(\eta_{LL},\,\eta_{RR},\,\eta_{RL}\right)$ correspond to
different CI models. The following CI scenarios with color-singlet
couplings between quarks are investigated:
\begin{center}
  \begin{tabular}{c|c}
    $\Lambda$ & $\left(\eta_{LL},\,\eta_{RR},\,\eta_{RL}\right)$\rbthm\\\hline
    $\Lambda^{\pm}_{LL}$    & $(\pm 1,\hphantom{\pm}0,\hphantom{\pm}0)\,$\rbtrr\\
    $\Lambda^{\pm}_{RR}$    & $(\hphantom{\pm}0,\pm 1,\hphantom{\pm}0)\,$\rbtrr\\
    $\Lambda^{\pm}_{VV}$    & $(\pm 1,\pm 1,\pm 1)\,$\rbtrr\\
    $\Lambda^{\pm}_{AA}$    & $(\pm 1,\pm 1,\mp 1)\,$\rbtrr\\
    $\Lambda^{\pm}_{(V-A)}$ & $(\hphantom{\pm}0,\hphantom{\pm}0,\pm 1)\,$\rbtrr
  \end{tabular}
\end{center}
Note that the models with positive (negative) $\eta_{LL}$ or $\eta_{RR}$ lead to
destructive (constructive) interference with the QCD terms and a lower
(higher) cross section in the limit of high partonic c.m. energies.
In all CI models discussed in this Letter, next-to-leading-order (NLO)  QCD corrections are
employed to calculate the cross sections. In proton-proton collisions the $\Lambda^{\pm}_{LL}$ and
$\Lambda^{\pm}_{RR}$ models result in identical tree-level cross sections and NLO corrections,
and consequently lead to the same sensitivity. For
$\Lambda^{\pm}_{VV}$ and $\Lambda^{\pm}_{AA}$, as well as for
$\Lambda^{\pm}_{(V-A)}$, the CI predictions are
identical at tree-level, but exhibit different NLO corrections and yield different sensitivity.

Measurements of dijet angular distributions at the Fermilab Tevatron have been
reported by the CDF~\cite{Abe:1996mj}
and D0~\cite{Abazov:2009mh,Abbott:2000kp} Collaborations, and at the LHC by the
CMS~\cite{CMS-PAPERS-EXO-11-017,CMS-PAPERS-QCD-10-016,CMS-PAPERS-EXO-10-002}
and ATLAS~\cite{Aad:2012pu,Aad:2011aj} Collaborations.
The most stringent limits to date on CI models calculated at tree-level
have been obtained by the CMS Collaboration
from the inclusive jet \pt spectrum~\cite{CMS-PAPERS-EXO-11-010}, which excludes
$\Lambda^+_{LL} < 9.9$\TeV and $\Lambda^-_{LL} < 14.3$\TeV.
Constraints on CI models with NLO corrections have been previously obtained from a
search in the dijet angular
distributions~\cite{CMS-PAPERS-EXO-11-017}, excluding in particular
$\Lambda^+_{LL} < 7.5$\TeV and $\Lambda^-_{LL} < 10.5$\TeV.

Dijet angular distributions are also sensitive to signatures from the
Arkani-Hamed--Dimopoulos--Dvali (ADD)
model~\cite{ArkaniHamed:1998nn,ArkaniHamed:1998rs}
of compactified extra dimensions (EDs) that provides a possible
solution to the hierarchy problem of the standard model (SM).
In the ADD model,
gravity is assumed to propagate in the entire higher-dimensional
space, while SM particles are confined to a (3+1) dimensional subspace.
As a result, the fundamental Planck scale \MD in the ADD model is
much smaller than the (3+1) dimensional Planck energy scale \Mpl,
which may lead to phenomenological effects that can be tested with
proton-proton collisions at the LHC.
The coupling of the graviton in higher-dimensional space to the SM
fields can be described by a (3+1)-dimensional tower of Kaluza--Klein
(KK) graviton excitations, each coupled to the energy-momentum tensor
of the SM field with gravitational strength. The effects of a
virtual graviton exchange can therefore be approximated at leading-order (LO) by an
effective (3+1)-dimensional theory that sums over KK excitations of a
virtual graviton. This sum is divergent, and therefore has to be
truncated at a certain energy scale of order \MD, where the effective theory is expected to break down.
Such a theory predicts a non-resonant enhancement of dijet production, whose angular
distribution differs from the QCD prediction.  Two parameterizations for
virtual graviton exchange in the ADD model are considered, namely the
Giudice--Rattazzi--Wells (GRW)~\cite{GRW} and the Han--Lykken--Zhang
(HLZ)~\cite{HLZ} conventions. Though not considered in this paper,
another convention by Hewett~\cite{Hewett} exists.
In the GRW convention the sum over the KK states is regulated by a
single cutoff parameter $\Lambda_T$. The HLZ convention describes the effective
theory in terms of two parameters, the cutoff scale $M_S$ and the number
of extra spatial dimensions \nED.
The parameters $M_{S}$ and \nED can be directly related to
$\Lambda_{T}$~\cite{landsberg}.  We consider scenarios with 2 to 6
EDs.
The case of $\nED=1$ is not considered since it would require an
ED of the size of the order of the solar system; the gravitational potential
at these distances would be noticeably modified and this case is therefore excluded.
The case of $\nED=2$ is special in
the sense that the relation between $M_{S}$ and $\Lambda_{T}$ also
depends on the parton-parton c.m. energy $\sqrt{\hat{s}}$.
Signatures from virtual graviton exchange have previously been sought in
dilepton~\cite{newCMSdilepton,newATLASdilepton},
diphoton~\cite{CMS-PAPERS-EXO-11-038,Aad:2012cy}, and
dijet~\cite{Abazov:2009mh,ATLASdijet,Franceschini:2011wr} final states, where the
most stringent limits come from the dilepton searches and range from 3.5 to 4.9\TeV.

In this Letter, we extend previous searches for contact
interactions to higher CI scales, for a wide range of models that
include the exact NLO QCD corrections to dijet production. In addition, we explore
various models of compactified extra dimensions. Using a data sample corresponding to an integrated luminosity of
\intlumi at $\sqrt s$ = 8\TeV, the measured dijet angular distributions, unfolded for
detector effects, are compared to QCD predictions at NLO, including
for the first time electroweak (EW) corrections.

\section{Event selection}

A detailed description of the CMS detector,
together with a definition of the coordinate
systems used and the relevant kinematic variables,
can be found in Ref.~\cite{bib_CMS}.
The central feature of the CMS apparatus is a superconducting solenoid of
6\unit{m} internal diameter, providing an axial field of
3.8\unit{T}. Within the solenoid are the silicon pixel and strip
trackers, which cover the region of pseudorapidity $\abs{\eta} < 2.5$, and
the lead tungstate crystal electromagnetic and the
brass and scintillator hadronic calorimeters, which surround the tracking volume and
cover $\abs{\eta} < 3$.
Muons are measured in gas-ionization detectors embedded
in the steel flux-return yoke of the solenoid with a coverage of $\abs{\eta} <
2.4$.

Events are reconstructed using a particle-flow
technique~\cite{CMS-PAS-PFT-09-001,CMS-PAS-PFT-10-001} which combines
information from all CMS subdetectors to identify and reconstruct in
an optimal way the
individual particle candidates (charged hadrons, neutral hadrons,
electrons, muons, and photons) in each event. These particle
candidates are clustered into jets using the anti-\kt
algorithm~\cite{Cacciari:2008gp} as implemented in the \fastjet
package~\cite{Cacciari:2011ma} with a size parameter $R = 0.5$. Jet
energy scale corrections~\cite{bib_jecjinst} derived from data and
Monte Carlo (MC) simulation are applied
to account for the response function of the calorimeters for hadronic showers.

The CMS trigger system uses a two-tiered system comprising a level-1 trigger
(L1) and a high-level trigger (HLT) to select physics events of
interest for further analysis. The selection criteria used in this analysis
are the inclusive single-jet triggers, which require one L1 jet and one
HLT jet with various thresholds on the jet \pt, as well as trigger
paths with thresholds on the dijet mass and scalar sum of the jet
\pt.
The \pt of jets is
corrected for the response of
the detector at both L1 and the HLT\@. The efficiency of each single-jet trigger is measured as a function of dijet mass \mjj
using events selected by a lower-threshold trigger.

Events with at least two reconstructed jets are selected from an
inclusive
jet sample and the two highest-\pt jets are used to measure the dijet
angular distributions for different ranges in \mjj.
In units of \TeV the \mjj ranges are (1.9, 2.4),
(2.4, 3.0), (3.0, 3.6), (3.6, 4.2), and $>$4.2. The lowest
\mjj range is chosen such that the trigger efficiency exceeds
99\% in all bins of \chijj considered in this analysis.
The two highest \mjj ranges were chosen to maximize the expected
sensitivity to the new physics signals considered.
Events with spurious jets from noise and noncollision
backgrounds are rejected by applying loose quality criteria~\cite{CMS-PAS-JME-10-001} to jet
properties and requiring a reconstructed primary vertex within $\pm$24\unit{cm} of the detector center along the beam line and within
2\unit{cm} of the detector center in the plane transverse to the
beam.
The main primary vertex is defined as the one with the largest
summed $\pt^2$ of its associated tracks.
The phase space for this analysis is defined by selecting
events with $\chijj < 16$ and $\yboost < 1.11$, where
$\yboost = \frac{1}{2}\abs{y_1 + y_2}$.
This choice of values restricts the two jets within $\abs{y}<2.5$.
The highest value of \mjj observed
in this data sample is 5.2\TeV.

\section{Cross section unfolding and uncertainties}

The measured \chijj distributions, defined as
$(1/\sigma_\text{dijet})(\rd\sigma_\text{dijet}/\rd\chi_\text{dijet})$, are corrected
for migration effects due to the finite jet energy and position resolutions of the
detector.  Fluctuations in the jet response cause event migrations in
\chijj as well as in dijet mass. Therefore, a two-dimensional unfolding
in these variables is performed using the D'Agostini
method~\cite{D'Agostini:1994zf} as implemented in the \RooUnfold
package~\cite{Adye:2011gm}. The unfolding corrections are determined
from a response matrix that maps the true \mjj and \chijj
distributions onto the measured ones.
This matrix is derived using particle-level jets from \HERWIGpp
version 2.5.0~\cite{herwig25,Bahr:2008pv} with the tune of version 2.4.
The jets are smeared in \pt with a double-sided Crystal-Ball
parameterization~\cite{oreglia} of the response, which takes into account the full jet
energy response including non-Gaussian tails. The unfolding
correction factors as a function of \chijj vary from less than 3\% in the lowest \mjj range to
less than 20\% in the highest \mjj range.

The main experimental systematic uncertainties in this analysis are
caused by the jet energy scale, the jet energy resolution, and
the unfolding modeling and detector simulation.
The overall jet energy scale uncertainty
varies between 1\% and 2\% and has a dependence on pseudorapidity of
less than 1\% per unit of $\eta$~\cite{bib_jecjinst}.
The jet energy scale uncertainty is divided into 21 uncorrelated sources~\cite{Collaboration:2014dp}.
The effect of each source is propagated to the dijet angular distributions
and then summed in quadrature to take into account uncorrelated
\pt- and $\eta$-dependent sources that could cancel if
varied simultaneously. The resulting
uncertainty in the \chijj distributions due to the jet energy
scale uncertainties is found to be less than 2.0\% (2.6\%) at
low (high) \mjj over all \chijj bins, and the maximum uncertainty in a given \mjj bin is typically
found to be in the lowest \chijj bin.

The jet energy resolution is known to within 10\% of its value in the phase space
considered in this analysis~\cite{bib_jecjinst}. The systematic uncertainty
in the \chijj distributions due to this effect was evaluated by
varying the width of the Gaussian core of the Crystal-Ball
parameterization of the response by $\pm$10\% and comparing the
resultant unfolding corrections before and after these changes. The
resulting uncertainty in the \chijj distributions is 0.5\% (1.5\%) in
the lowest (highest) \mjj range.  In addition, a systematic
uncertainty in the tails of the jet response function is evaluated by
determining a correction factor using a Gaussian
ansatz~\cite{bib_jecjinst} rather than the double-sided Crystal-Ball
(Gaussian with tails) function to parameterize the response.
Since the Gaussian assumption corresponds to the extreme case of the complete
absence of a tail, the associated uncertainty has been taken to be 50\% of the difference
between this correction and the nominal correction based on the
Crystal-Ball function. This covers the uncertainty in the
understanding of the tails from jet resolution tail measurements.  The
size of this uncertainty varies from less than 1\% in the lowest \mjj
range to less than 13\% in the highest \mjj range.

A systematic uncertainty in the unfolding due to the use of a parameterized
model of the jet \pt and position resolutions to determine the
unfolding correction factors is estimated by comparing the
smeared \chijj distributions to the ones from a detailed simulation of
the CMS detector using \GEANTfour~\cite{Agostinelli:2002hh}. This
uncertainty is found to be less than 0.4\% (5\%) in the lowest
(highest) \mjj range.  A further systematic uncertainty in the unfolding
for the modeling of the dijet spectra with \HERWIGpp [0.1\% (1.2\%)
in the lowest (highest) \mjj range], is estimated from a comparison of the
unfolding corrections from \HERWIGpp with those obtained
from \PYTHIAE version~8.165~\cite{Sjostrand:2007gs} with tune 4C~\cite{tune4c}.

The uncertainty from additional interactions in the same
proton bunch crossing as the interaction of interest, called pileup,
is determined in simulation by varying the minimum bias
cross section within its measured uncertainty of 6\%~\cite{minbiasxsec}.
No significant effect is observed. Though in the statistical analysis
of the data the uncertainties are treated separately, for display in
tables and figures, the total experimental systematic uncertainty in the
\chijj distributions is calculated as the quadratic sum of the
contributions due to the uncertainties in the jet energy calibration,
jet \pt resolution, and unfolding correction. The total uncertainty including
statistical uncertainties is less than 2.5\% (49\%) for the lowest (highest) \mjj range.
Experimental uncertainties are evaluated for both the QCD background
and signal predictions, however,
the resulting uncertainties do not differ significantly.

\section{Theoretical predictions}

The normalized dijet angular distributions are compared to
the predictions of perturbative QCD\@. The NLO calculation is provided by
\NLOJETPP version~4.1.3~\cite{Nagy:2001fj,Nagy:2003tz} within the
\fastNLO framework version~2~\cite{Kluge:2006xs,Britzger:2012bs}.
The factorization (\muf) and renormalization (\mur) scales
are defined to be the average \pt of the two jets, \avept.
Electroweak corrections for dijet production have been derived in
Ref.~\cite{Dittmaier:2012kx}, the authors of which provided us with the
corresponding corrections for the \chijj distributions.
These corrections change the predictions
of the normalized \chijj distributions by up to 4\% (14\%) at low
(high) \mjj.
Since fast re-evaluation techniques for different choices of PDFs or scales
are not yet available for the electroweak correction part of the theory,
the factors have been applied here without additional uncertainties.
A figure showing these corrections can be found in the Appendix.
The impact of non-perturbative effects such as hadronization and multiple
parton interactions is estimated using \PYTHIAE and \HERWIGpp. These
effects are found to be negligible.

The dominant uncertainty in the QCD predictions is associated with the
choice of the \mur and \muf scales and is evaluated following the
proposal in Ref.~\cite{Banfi:2010xy} by varying the default choice of
scales in the following six combinations: $(\muf/\avept$,
$\mur/\avept) = (1$/2, 1/2), (1/2, 1), (1,1/2), (2, 2), (2,
1), and (1, 2). These scale variations change the QCD predictions
of the normalized \chijj distributions by less than 9\% (18\%) at low
(high) \mjj. The uncertainty due to the choice of parton distribution functions (PDF) is determined
from the 22 uncertainty eigenvectors of CT10~\cite{Lai:2010vv} using the procedure
described in Ref.~\cite{Lai:2010vv}, and is found to be less than
0.6\% (1.0\%) at low (high) \mjj. A summary of the systematic
uncertainties in the theoretical predictions is given in
Table~\ref{tab:sys} together with the experimental ones. In the
highest \mjj range, the dominant experimental contribution
is the statistical uncertainty while the dominant theoretical
contribution is the QCD scale uncertainty.

\begin{table*}[h!tb]
  \begin{center}
    \topcaption{Summary of the experimental and theoretical
      uncertainties in the normalized \chijj distributions. For the
      lowest, second highest
      and highest \mjj ranges, the relative shift (in \%) of the
      lowest \chijj bin from its nominal value is quoted. While in
      the statistical analysis each systematic uncertainty is represented by a
      change of the \chijj distribution correlated among all \chijj
      bins, this table summarizes each uncertainty by a representative
      number to demonstrate the relative contributions.}
    \label{tab:sys}
    \begin{tabular}{l|c|c|c}
      \hline
      \multirow{2}{*}{\centering Uncertainty} & $1.9<\mjj<2.4$ & $3.6<\mjj<4.2$ & $\mjj>4.2$ \\
                                              &    \TeV(\%)  &  \TeV(\%)   &  \TeV(\%)       \\
      \hline
      Statistical & 1.0 & 2.3 & 47 \\
      Jet energy scale  & 2.0 & 2.1 & 2.5 \\
      Jet energy resolution (tails) & 1.0 & 2.0 & 13 \\
      Jet energy resolution (core) & 0.5 & 0.6 & 1.5 \\
      Unfolding, modeling & 0.1 & 1.2 & 1.2 \\
      Unfolding, detector simulation & 0.4 & 1.0 & 5.0 \\
      Pileup & $<$0.1 & $<$0.1 & $<$1.0 \\
      \hline
      Total experimental & 2.5 & 4.1 & 49 \\
      \hline
      QCD NLO scale (6 variations of $\mu_R$ and $\mu_F$) &
      $_{-3.4}^{+9.0}$ & $_{-4.0}^{+11}$ & $_{-6.3}^{+18}$ \\
      PDF (CT10 eigenvectors) & 0.6 & 0.7 & 1.0 \\
      Non-perturbative effects  & $<$1.0 & $<$1.0 & $<$0.2 \\
      \hline
      Total theoretical & 9 & 11 & 18 \\
      \hline
    \end{tabular}
  \end{center}
\end{table*}
For calculating the CI terms as well as the interference between the
CI terms and QCD terms at LO and NLO in QCD the \CIJET program version 1.0~\cite{Gao:2011ha} has been employed.
The CI models at LO
are cross-checked with the implementation in \PYTHIAE and found to be
consistent. The ADD predictions are calculated with \PYTHIAE.

\section{Results}

\begin{figure}[h!!!]
  \centering
  \includegraphics[width=\cmsFigWidth]{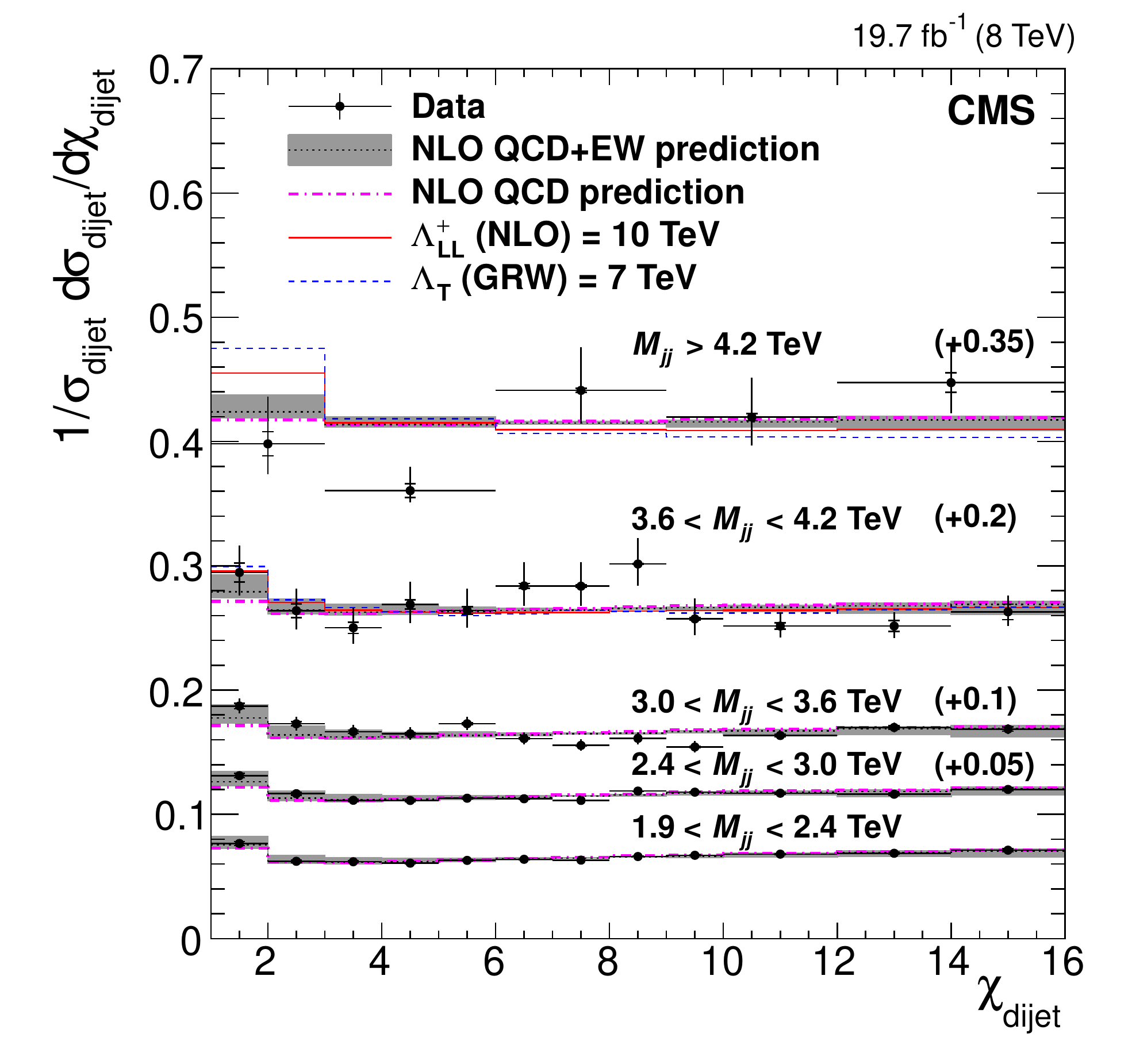}
  \caption{Normalized \chijj distributions for \intlumi of integrated
    luminosity at $\sqrt{s} = 8$\TeV. The corrected data distributions are compared to NLO
    predictions with EW corrections (black dotted line).  For clarity the
    individual distributions are shifted vertically by offsets
    indicated in parentheses. Theoretical uncertainties are indicated as
    a gray band. The error bars represent statistical and experimental
    systematic uncertainties combined in quadrature. The ticks on the
    error bars represent experimental systematic uncertainties
    only. The horizontal bars indicate the bin widths.
    The NLO QCD prediction without EW corrections is also shown (purple dashed dotted).
    The prediction for SM+CI with
    $\Lambda_{LL}^{+}$~(NLO)~$=10$\TeV is shown (red solid line), and
    so is the prediction for SM+ADD with $\Lambda_{T}$~(GRW)~$=7$\TeV
    (blue dashed line).}
  \label{fig:data_results}
\end{figure}

In Fig.~\ref{fig:data_results} the measured \chijj distributions, corrected for instrumental effects and
normalized by their respective event counts, for all \mjj ranges, are compared to
theoretical predictions. The data are
well described by NLO calculations that incorporate EW corrections.
No significant deviation from the SM predictions is observed.
The distributions are also compared to predictions for SM+CI with
$\Lambda_{LL}^{+}$~(NLO)~$=10$\TeV and predictions for SM+ADD with
$\Lambda_{T}$~(GRW)~$=7$\TeV.

The measured \chijj distributions are used to determine exclusion limits on CI models that
include full NLO QCD corrections to dijet production induced by
contact interactions calculated with \CIJET. Limits are also extracted for
CI models calculated at LO with \CIJET and
ADD models implemented in \PYTHIAE.
To take into account the NLO QCD and EW corrections in these LO
models, the cross section difference
$\sigma^{\text{QCD}}_{\text{NLO+EW corr}} - \sigma^{\text{QCD}}_{\text{LO}}$
is evaluated for each \mjj and \chijj bin and added to the \PYTHIAE+ADD and LO QCD+CI
predictions. With this procedure, an SM+CI (SM+ADD) prediction is
obtained where the QCD terms are corrected to NLO with EW corrections while the CI (ADD)
terms are calculated at LO.
The variations due to theoretical uncertainties associated with scales
and PDFs are applied only to the QCD terms of the prediction,
thereby treating the effective new physics terms as fixed benchmark terms.

\begin{figure}[htbp]
  \centering
  \includegraphics[width=0.48\textwidth]{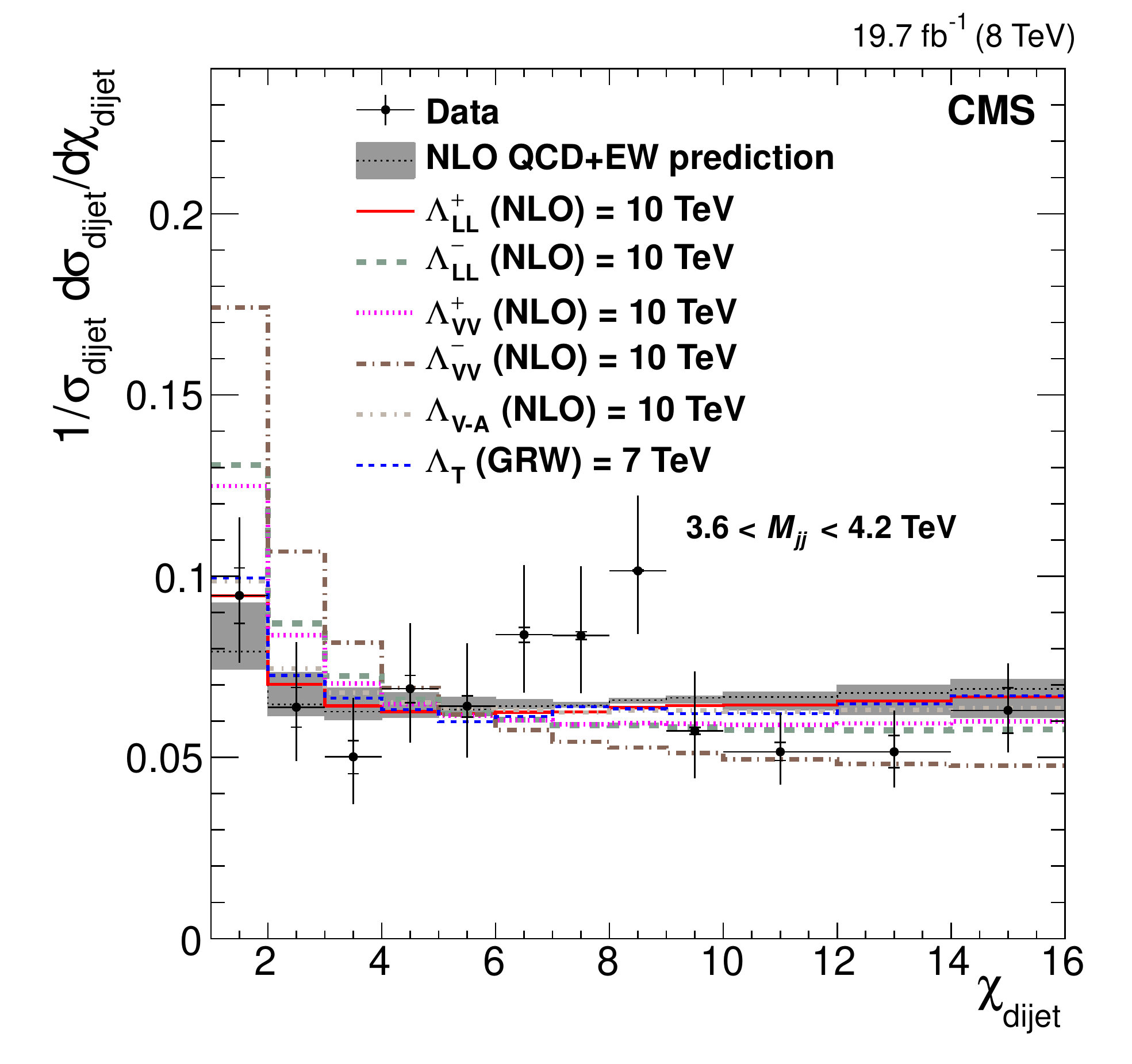}
  \includegraphics[width=0.48\textwidth]{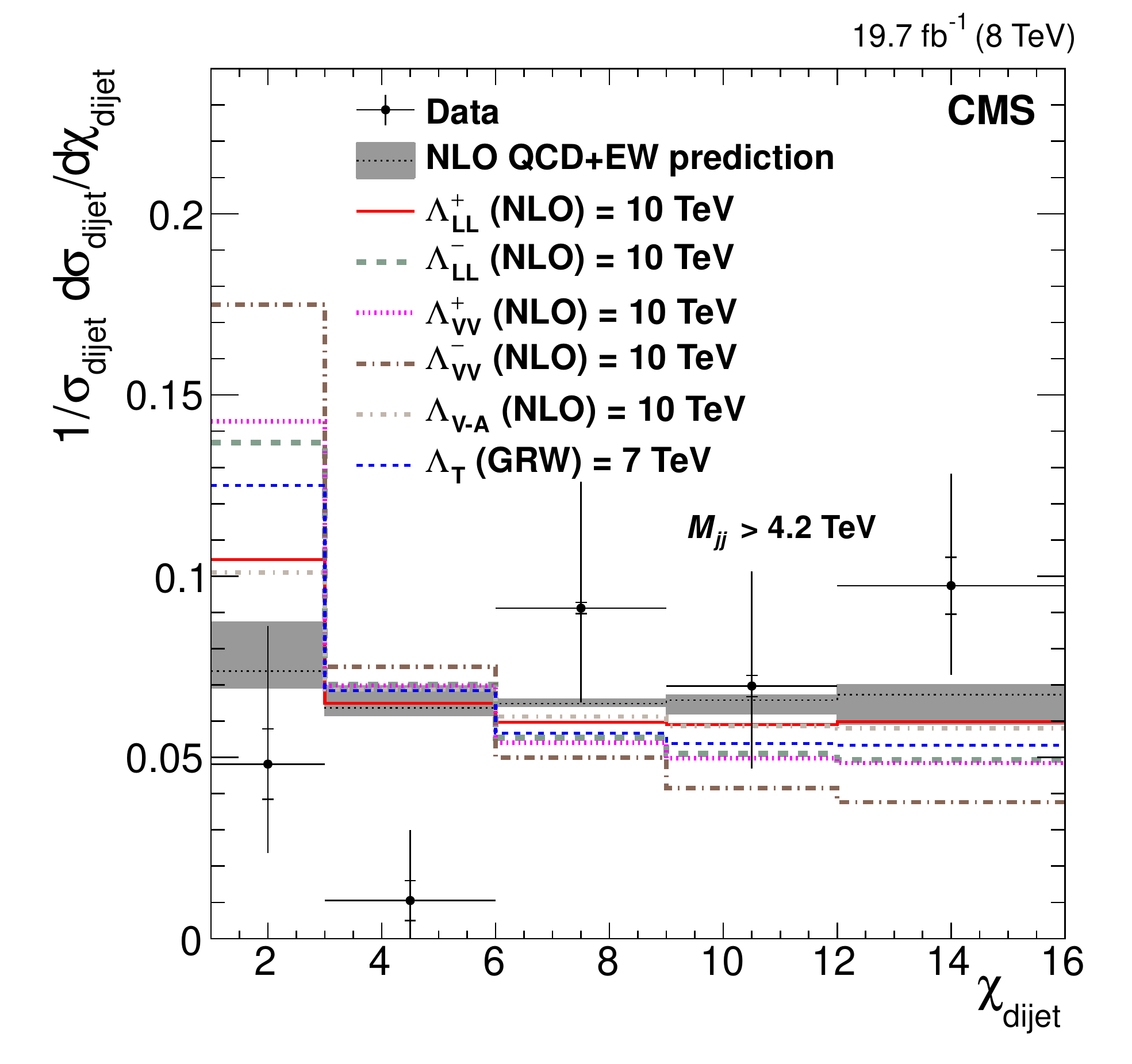}
  \caption{Normalized \chijj distributions in the two highest \mjj ranges. The corrected data
    distributions are compared to NLO predictions with EW corrections
    (black dotted line). Theoretical uncertainties are indicated as
    gray bands. The error bars represent statistical and experimental
    systematic uncertainties combined in quadrature. The ticks on the
    error bars represent experimental systematic uncertainties
    only. The horizontal bars indicate the bin widths. The predictions for the various CI and ADD models are
    overlaid.}
  \label{fig:data_results_theories}
\end{figure}

In Fig.~\ref{fig:data_results_theories}, the \chijj distributions for
the two highest \mjj ranges are compared to various CI and ADD models.
Only the two highest \mjj ranges are used to determine limits of
CI and ADD model parameters since the added sensitivity from the lower \mjj ranges is negligible.

We quantify the significance of a NP signal with respect to
the SM-only hypothesis by means of the likelihood for the SM-only,
$L_\mathrm{SM}$, and the likelihood for the SM with new physics, $L_\mathrm{SM+NP}$.
The $L_\mathrm{SM}$ and $L_\mathrm{SM+NP}$ are defined as products of Poisson likelihood functions
for each bin in \chijj for the two highest ranges of \mjj.
The predictions for each \mjj range are normalized to the number of
observed events in that range. The $p$-values for the two hypotheses,
$p_\mathrm{SM+NP}(q \geq q_\text{obs})$ and $p_\mathrm{SM}(q \leq
q_\text{obs})$, are based on the log-likelihood ratio $q = -2 \ln
(L_\mathrm{SM+NP}/L_\mathrm{SM})$. They are evaluated from
ensembles of pseudo-experiments, in which systematic uncertainties are
taken into account via nuisance parameters which affect the \chijj
distribution, varied within their Gaussian uncertainties when
generating the distributions of $q$~\cite{cousinshighland}.

We note that there is an observed difference between
the NLO QCD calculations with EW corrections
and
the NLO QCD-only hypothesis
in the above defined likelihood ratio,
which corresponds to a significance of 1.1 standard deviation.

The agreement of the data with the SM-only hypothesis is estimated by
calculating $p_\mathrm{SM}(q \leq q_\text{obs})$ for each \mjj bin separately.
The largest difference is found in the \mjj range 3.0--3.6\TeV with a
probability of 17\% to obtain a deviation from the SM-only hypothesis
larger than the observed, corresponding to a significance of 1.4 standard deviations.
Including the two highest \mjj bins in the likelihood reduces this significance
to 0.9 standard deviations, corresponding to a probability of 39\%.

A modified-frequentist
approach~\cite{junk:1999,read:2002,cousinshighland} is used to
set exclusion limits on the scale $\Lambda$.
Limits on
the SM+NP models are set based on the quantity $\mathrm{CL}_\mathrm{s} =
p_\mathrm{SM+NP}(q\geq q_\text{obs}) / (1-p_\mathrm{SM}(q \leq
q_\text{obs}))$, which is required to be 0.05 for a 95\%
confidence level (CL)  exclusion. The observed and expected
exclusion limits on different CI and ADD
models obtained in this analysis at 95\% CL are listed in
Tables~\ref{tab:limitsCI} and~\ref{tab:limitsADD} respectively. Note
that the CI predictions with exact NLO QCD corrections show a smaller
enhancement at low \chijj relative to QCD than do the corresponding LO
CI predictions, as described in detail in Ref.~\cite{Gao:2012qpa}, and
therefore result in less stringent limits.

\begin{table}[htbp]
  \centering
    \topcaption{Observed and expected exclusion limits at 95\%
      CL for various CI models. The
      uncertainties in the expected limits considering statistical
      and systematic effects for the SM-only hypothesis are also given.}
    \label{tab:limitsCI}
    \begin{tabular}{l|c|c}
     \hline
      \multirow{2}{*}
      {\centering Model} & Observed & Expected \\
                         & (\TeVns{})                & (\TeVns{}) \\
      \hline
      $\Lambda_{LL/RR}^{+}$ (LO) & 10.3 & 9.8 $\pm$ 1.0\\
      $\Lambda_{LL/RR}^{-}$ (LO) & 12.9 & 12.4 $\pm$ 2.2\\
      $\Lambda_{LL/RR}^{+}$ (NLO) & 9.0 & 8.7 $\pm$ 0.8\\
      $\Lambda_{LL/RR}^{-}$ (NLO) & 11.7 & 11.4 $\pm$ 1.8\\
      $\Lambda_{VV}^{+}$ (NLO) & 11.3 & 10.8 $\pm$ 1.1\\
      $\Lambda_{VV}^{-}$ (NLO) & 15.2 & 14.6 $\pm$ 2.6\\
      $\Lambda_{AA}^{+}$ (NLO) & 11.4 & 10.9 $\pm$ 1.1\\
      $\Lambda_{AA}^{-}$ (NLO) & 15.1 & 14.5 $\pm$ 2.6\\
      $\Lambda_{(V-A)}^{+}$ (NLO) & 8.8 & 8.5 $\pm$ 1.1\\
      $\Lambda_{(V-A)}^{-}$ (NLO) & 8.9 & 8.6 $\pm$ 1.2\\
      \hline
    \end{tabular}
\end{table}

\begin{table}[htb]
  \begin{center}
    \topcaption{Observed and expected exclusion limits at 95\%
      CL for various ADD models in LO. The
      uncertainties in the expected limits considering statistical
      and systematic effects for the SM-only hypothesis are also given.}
    \label{tab:limitsADD}
    \begin{tabular}{l|c|c}
      \hline
      \multirow{2}{*}
      {\centering Model} & Observed & Expected \\
                         & (\TeVns{})                & (\TeVns{}) \\
      \hline
      ADD $\Lambda_T$ (GRW)           & 7.1 & 6.8 $\pm$ 0.5 \\
      ADD $M_S$ (HLZ) $\nED=2$ & 6.9 & 6.6$\pm$ 0.4\\
      ADD $M_S$ (HLZ) $\nED=3$ & 8.4 & 8.0$\pm$ 0.6\\
      ADD $M_S$ (HLZ) $\nED=4$ & 7.1 & 6.8$\pm$ 0.5\\
      ADD $M_S$ (HLZ) $\nED=5$ & 6.4 & 6.1$\pm$ 0.5\\
      ADD $M_S$ (HLZ) $\nED=6$ & 5.9 & 5.7$\pm$ 0.4\\
     \hline
    \end{tabular}
  \end{center}
\end{table}

\begin{figure}[h!tbp]
  \centering
  \includegraphics[width=\cmsFigWidth]{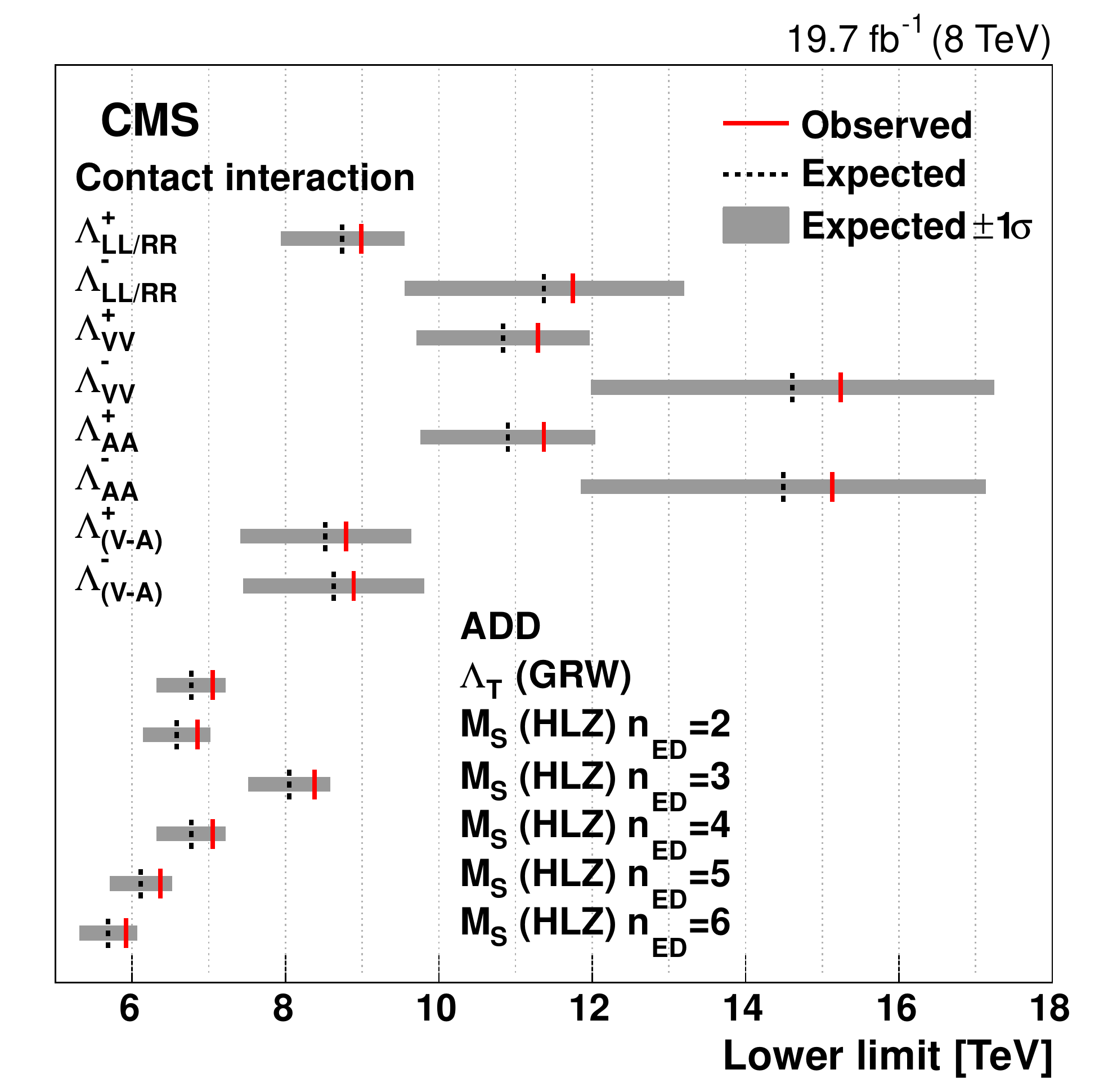}
  \caption{Observed (solid lines) and expected (dashed lines) 95\%
    CL lower limits for the CI scales
    $\Lambda$ for different compositeness models (NLO), for the ADD model scale with GRW
    parameterization $\Lambda_{T}$ and for the ADD model scale with HLZ parameterization
    $M_{S}$. The gray bands indicate the corresponding uncertainties in the expected
    exclusion limits.}
  \label{fig:limit_summaries}
\end{figure}

These results are also summarized in Fig.~\ref{fig:limit_summaries}.
The limits on $M_S$ for the different $\nED$
($\nED\ge 2$) directly follow from the limit for $\Lambda_T$. As
a cross check, the limits for the
CI scale $\Lambda_{LL/RR}^{+}$ are also determined for the
case in which the data are not corrected for detector effects and instead the simulation predictions are convoluted with the detector resolutions.
The extracted limits are found to agree with the quoted ones within 1\%.
We also quantify the effect of the inclusion of EW corrections in the QCD prediction on the $\Lambda_{LL/RR}^{+}$ (LO) observed limit,
which would be reduced from 10.3\TeV to 9.8\TeV if EW corrections were neglected.

\section{Summary}

Normalized dijet angular distributions have been measured
with the CMS detector over a wide range of dijet invariant masses.
No significant deviation from the standard model predictions is observed.
Lower limits are set on the contact interaction scale
for a variety of quark compositeness models that include NLO
QCD corrections and on the cutoff scale for the ADD models with extra dimensions.
The 95\% confidence level lower limits on the contact interaction scale
$\Lambda$ are in the range 8.8--15.2\TeV. The improved
description of the data resulting from the inclusion of the electroweak
corrections yields approximately 5\% higher limits.
The lower limits on the cutoff scales in the ADD models, $\Lambda_{T}$ (GRW) and $M_{S}$
(HLZ), are in the range 5.9--8.4\TeV. These results represent the
most stringent set of limits on contact interaction scale, modelled at NLO,
and the best limits on the benchmark ADD model to date.

\begin{acknowledgments}
We would like to thank S.~Dittmaier and A.~Huss for providing us with
the electroweak correction factors. We congratulate our colleagues in the CERN accelerator departments for the excellent performance of the LHC and thank the technical and administrative staffs at CERN and at other CMS institutes for their contributions to the success of the CMS effort. In addition, we gratefully acknowledge the computing centers and personnel of the Worldwide LHC Computing Grid for delivering so effectively the computing infrastructure essential to our analyses. Finally, we acknowledge the enduring support for the construction and operation of the LHC and the CMS detector provided by the following funding agencies: BMWFW and FWF (Austria); FNRS and FWO (Belgium); CNPq, CAPES, FAPERJ, and FAPESP (Brazil); MES (Bulgaria); CERN; CAS, MoST, and NSFC (China); COLCIENCIAS (Colombia); MSES and CSF (Croatia); RPF (Cyprus); MoER, ERC IUT and ERDF (Estonia); Academy of Finland, MEC, and HIP (Finland); CEA and CNRS/IN2P3 (France); BMBF, DFG, and HGF (Germany); GSRT (Greece); OTKA and NIH (Hungary); DAE and DST (India); IPM (Iran); SFI (Ireland); INFN (Italy); MSIP and NRF (Republic of Korea); LAS (Lithuania); MOE and UM (Malaysia); CINVESTAV, CONACYT, SEP, and UASLP-FAI (Mexico); MBIE (New Zealand); PAEC (Pakistan); MSHE and NSC (Poland); FCT (Portugal); JINR (Dubna); MON, RosAtom, RAS and RFBR (Russia); MESTD (Serbia); SEIDI and CPAN (Spain); Swiss Funding Agencies (Switzerland); MST (Taipei); ThEPCenter, IPST, STAR and NSTDA (Thailand); TUBITAK and TAEK (Turkey); NASU and SFFR (Ukraine); STFC (United Kingdom); DOE and NSF (USA).
\end{acknowledgments}

\bibliography{auto_generated}   % will be created by the tdr script.
\appendix
\section{EW corrections to dijet angular distributions}
\begin{figure}[htb]
  \centering
  \includegraphics[width=\cmsFigWidth]{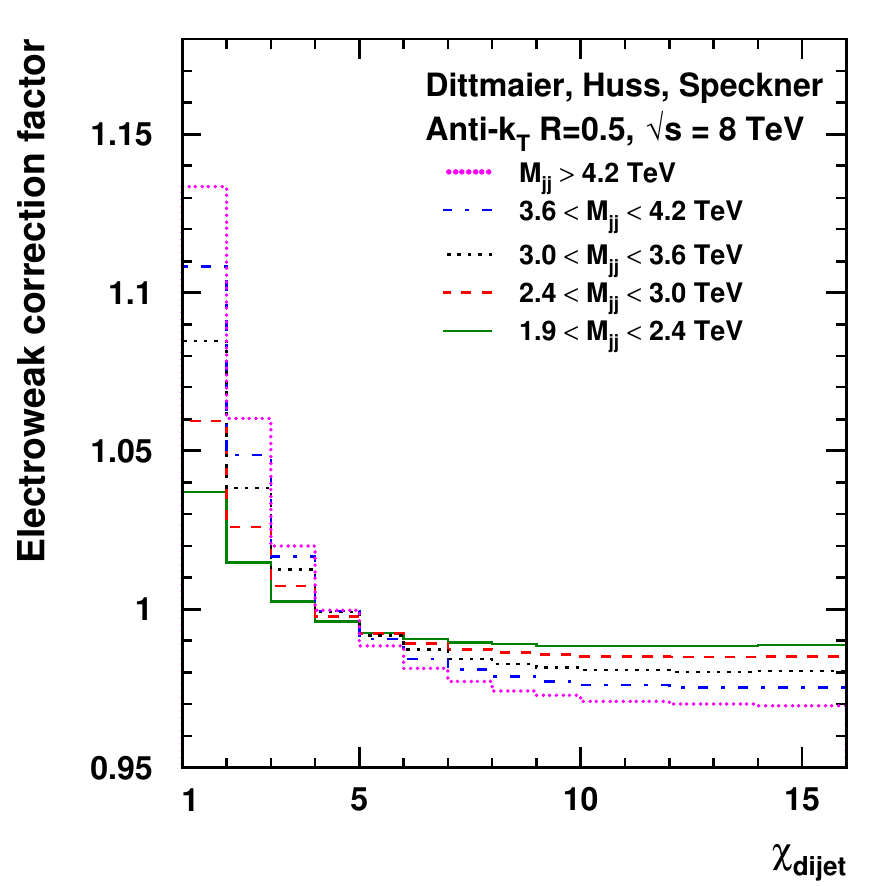}
  \caption{Electroweak correction factors versus \chijj for each \mjj
    range, derived by the authors of Ref.~\cite{Dittmaier:2012kx} at
    8\TeV c.m. energy with \avept as choice for the \mur and \muf
    scales and the CT10-NLO PDF set.}
  \label{fig:ewk}
\end{figure}

Figure~\ref{fig:ewk} shows the EW corrections to the dijet angular
distributions. The corrections are based on the same calculations and
tools used to derive the EW corrections to inclusive jet and dijet
production cross sections published in Ref.~\cite{Dittmaier:2012kx}.
The authors of Ref.~\cite{Dittmaier:2012kx} have provided the
exact numbers to be applied to the dijet angular distribution as
presented in this paper. The EW corrections change the predictions of
the normalized \chijj distributions by up to 4\% (14\%) at low (high)
\mjj.

\cleardoublepage \section{The CMS Collaboration \label{app:collab}}\begin{sloppypar}\hyphenpenalty=5000\widowpenalty=500\clubpenalty=5000\input{EXO-12-050-authorlist.tex}\end{sloppypar}
\end{document}

%% file: EXO-12-050-authorlist.tex
\textbf{Yerevan Physics Institute,  Yerevan,  Armenia}\\*[0pt]
V.~Khachatryan, A.M.~Sirunyan, A.~Tumasyan
\vskip\cmsinstskip
\textbf{Institut f\"{u}r Hochenergiephysik der OeAW,  Wien,  Austria}\\*[0pt]
W.~Adam, T.~Bergauer, M.~Dragicevic, J.~Er\"{o}, M.~Friedl, R.~Fr\"{u}hwirth\cmsAuthorMark{1}, V.M.~Ghete, C.~Hartl, N.~H\"{o}rmann, J.~Hrubec, M.~Jeitler\cmsAuthorMark{1}, W.~Kiesenhofer, V.~Kn\"{u}nz, M.~Krammer\cmsAuthorMark{1}, I.~Kr\"{a}tschmer, D.~Liko, I.~Mikulec, D.~Rabady\cmsAuthorMark{2}, B.~Rahbaran, H.~Rohringer, R.~Sch\"{o}fbeck, J.~Strauss, W.~Treberer-Treberspurg, W.~Waltenberger, C.-E.~Wulz\cmsAuthorMark{1}
% ...
\vskip\cmsinstskip
% ...
1:~~Also at Vienna University of Technology, Vienna, Austria\\
2:~~Also at CERN, European Organization for Nuclear Research, Geneva, Switzerland\\
% ...
